\pdfoutput=1

\PassOptionsToPackage{pdfpagelabels=false}{hyperref} 

\documentclass[useAMS]{mnras}

\let\bibhang\relax
\usepackage{savesym}
\usepackage{natbib}
\usepackage{graphicx}
\usepackage{gensymb}
\usepackage{amsmath}
\usepackage{multirow}

\begin{document}
\title[X-ray observations of Lanning 386]{\textit{XMM-Newton} Observations of the Peculiar Cataclysmic Variable Lanning 386: X-ray Evidence for a Magnetic Primary}
\author[M. R. Kennedy et al.]
{M. R. Kennedy$^{1,2}$\thanks{Email: markkennedy@umail.ucc.ie}, P. Callanan$^1$, P. M. Garnavich$^2$, M. Fausnaugh$^3$, J. C. Zinn$^3$ \\$^1$Department of Physics, University College Cork, Ireland\\$^2$Department of Physics, University of Notre Dame, Notre Dame, IN 46556\\$^3$Department of Astronomy, Ohio State University, Columbus, OH 43210}
\maketitle
\date{}
\pagerange{\pageref{firstpage}--\pageref{lastpage}} \pubyear{2016}
\label{firstpage}

\begin{abstract}
We present the first X-ray observations of the eclipsing cataclysmic variables Lanning 386 and MASTER OTJ192328.22+612413.5,  possible SW Sextantis systems. The X-ray light curve of Lanning 386 shows deep eclipses, similar to the eclipses seen in the optical light curve, confirming the high inclination of the system. There is evidence of a periodicity between 17-22 min in the X-ray and optical light curves of Lanning 386, which is associated with quasi-periodic oscillations. This system also displays a hard X-ray spectrum which is well fit by a partially covered, absorbed 2 temperature plasma. The cool plasma temperature (0.24$^{+0.17}_{-0.08}$ keV) and hot plasma temperature (9$^{+4}_{-2}$ keV) are not atypical plasma temperatures of known intermediate polar systems. Based on this model, we suggest that Lanning 386 is an intermediate polar with a high accretion rate. The hot plasma temperature limits the white dwarf mass to $>$0.5 M$_{\odot}$. From the optical spectrum obtained using the Large Binocular Telescope, we find  that the secondary in the system is consistent with an M5V star, and refine the distance to Lanning 386 to be 160$\pm$50 pc. Finally, we use the high time resolution of the optical spectra to crudely constrain the magnetic moment of the white dwarf in Lanning 386. J1923 was not detected during the observations, but the upper limit on the flux is inline with J1923 and Lanning 386 being related.
\end{abstract}


\section{Introduction}
Cataclysmic variables (CVs) are close binary systems where a white dwarf primary accretes material from a late-type main sequence companion which is filling its Roche lobe. All CVs produce X-rays, but the X-ray emitting region depends on the magnetic field strength of the white dwarf. In non-magnetic systems, an accretion disk forms around the white dwarf and hard X-rays are thought to be produced between the inner radius of the accretion disk and the surface of the white dwarf. In strongly magnetic systems (AM Her systems, or polars), there is no accretion disk and material follows magnetic fields lines from the L1 point to the white dwarf. Soft X-rays in polars come from the region where this accretion stream encounters the shock region above the surface of the white dwarf \citep{beuermann2004}. In weakly magnetic systems (DQ Herculis stars, or intermediate polars; IPs), an accretion disk forms up until the magnetic pressure of the white dwarf overcomes the ram pressure of the disk, where the material starts to follow accretion curtains to the poles of the WD. In IPs, X-rays are thought to be generated in the accretion column, which is the region where the infalling material is shocked. This narrow region lies between the top of the accretion curtain and the surface of the white dwarf\citep{Patterson1994}. For an overview of the X-ray properties of cataclysmic variables, see \cite{2012MmSAI..83..585B}.

Lanning 386 and MASTER OTJ192328.22+612413.5 (hereafter J1923) are 2 cataclysmic variables that avoid easy classification. Lanning 386 has an orbital period of 3.94 hours \citep{Brady2008} and J1923 has an orbital period of 4.02 hours \citep{kennedy16}. Both systems show low amplitude, frequent outbursts with $\delta m \approx 2$, and display quasi-periodic oscillations (QPOs) in their optical light curves during quiescence, which are thought to arise in systems with magnetic primaries that are undergoing very high accretion rates (\citealt{Patterson2002}; \citealt{Warner2004}). The light curves of Lanning 386 and J1923 show deep eclipses, and the inclinations of these systems are thought to be high \citep{kennedy16}. The spectra of Lanning 386 shows strong  H and \ion{He}{i} lines in quiescence and outburst. It also shows \ion{He}{ii} and \ion{C}{IV} lines and the Bowen Blend complex in outburst. J1923 shows slightly double peaked H lines in quiescence. 

It has been suggested that the primaries in these systems may be magnetic, and the accretion disks might be truncated close to the surface of the white dwarf, leading to the formation of an accretion curtain \citep{kennedy16}. Here, we present the first X-ray observations of Lanning 386, taken using the \textit{XMM-Newton} telescope, in an attempt to constrain the magnetic nature of this peculiar CV. J1923 was also observed by \textit{XMM-Newton} as part of the same program due to the similarities between Lanning 386 and J1923 in their optical light curves and spectra in quiescence. However, only upper limits on the X-ray flux for J1923 are presented here.

\section{Observations} \label{ObsSect}
Lanning 386 was observed four times by \textit{XMM-Newton}, while J1923 was observed once (see Table \ref{xmm_observations} for full details). The EPIC-pn \citep{2001A&A...365L..18S} and EPIC-MOS \citep{2001A&A...365L..27T} instruments were operated in Large Frame mode with thin filters inserted. The RGS spectrographs (\citealt{1998sxmm.confE...2B}; \citealt{2001A&A...365L...7D}) were operated in their normal spectroscopy modes. The optical monitor (OM; \citealt{2001A&A...365L..36M}) was operated in fast timing mode with a V band filter inserted.

\begin{table*}
	\centering
	\caption{Observations of Lanning 386 and J1923 taken by \textit{XMM-Newton}}
	\begin{tabular}{r c c c c c c c c}
		\hline
		Object			& Obs No.	& Rev	& Start Date		& Instrument	& Filter	& Magnitude	& EXP Time 	& Total GTI\\
		\hline\hline
		Lanning 386 	& 1			& 2816	& 2015-04-26		& EPIC-pn		& Thin		& 			& 14.7 ks	& 13.4 ks\\
						&			&		&					& EPIC-MOS		& Thin		& 			& 16.7 ks\\
						&			&		&					& OM			& V			& 16.81		& 15.4 ks\\
						& 2			& 2825	& 2015-05-13		& EPIC-pn		& Thin		& 			& 20.2 ks 	& 13 ks\\
						&			&		&					& EPIC-MOS		& Thin		& 			& 22.2 ks\\
						&			&		&					& OM			& V			& 16.86		& 18.8 ks\\
						& 3			& 2828	& 2015-05-20		& EPIC-pn		& Thin		& 			& 16.7 ks 	& 8.9 ks\\
						&			&		&					& EPIC-MOS		& Thin		& 			& 18.7 ks\\
						&			&		&					& OM			& V			& 16.88		& 9.4 ks\\
						& 4			& 2829	& 2015-05-21		& EPIC-pn		& Thin		& 			& 14.7 ks 	& 14.7 ks\\
						&			&		&					& EPIC-MOS		& Thin		& 			& 16.7 ks\\
						&			&		&					& OM			& V			& 16.85		& 11 ks\\
		J1923			& 5			& 2809	& 2015-04-12		& EPIC-pn		& Thin		& 			& 26.4 ks 	& 12.4 ks\\
						&			&		&					& EPIC-MOS		& Thin		& 			& 26.9 ks\\
						&			&		&					& OM			& V			& $>$17.96	& 26.4 ks\\
		\hline
	\end{tabular}
	\label{xmm_observations}
\end{table*}

The four observations of Lanning 386 were scheduled randomly, in order to maximise the chance of observing it in outburst, as Lanning 386 spends approximately 25\% of its time in outburst \citep{Brady2008}.

\begin{figure}
	\includegraphics[width=80mm]{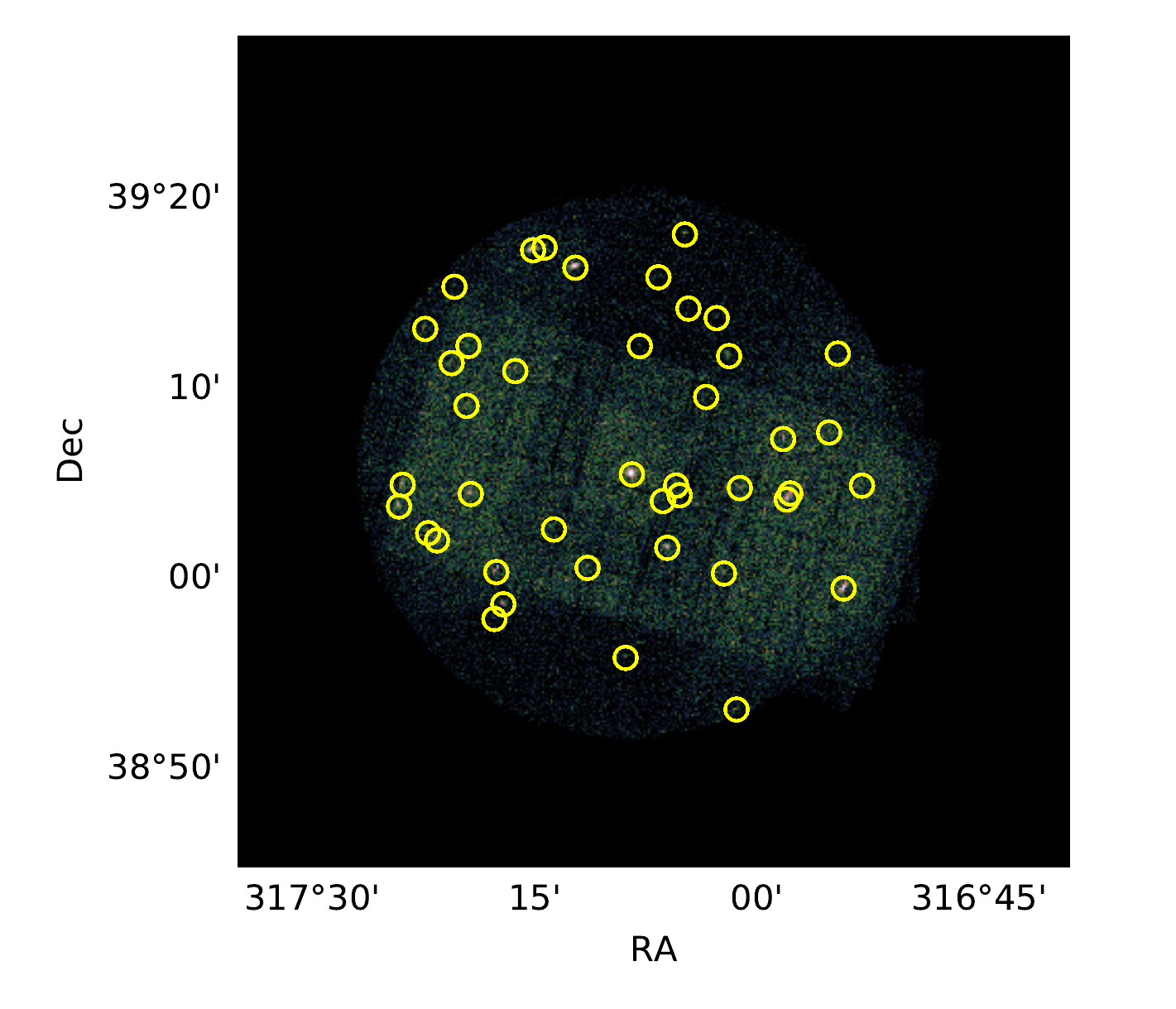}
	\caption{The combined images of Lanning 386 from observations 1, 2, 3 and 4 using the EPIC-pn and -MOS instruments. The yellow circles are all detected sources using {\sc edetect\_chain}. Lanning 386 is in the centre of the frame.}
	\label{lanxrayimg}
\end{figure}

The data were reduced and analysed using the \textit{XMM-Newton} Science Analysis Software ({\sc SAS} v14.0.0; \citealt{XMMHandbook}). The data from the EPIC instruments were processed using the {\sc emproc} and {\sc epproc} tasks. The {\sc edetect\_chain} command was used for source detection. The EPIC-pn and -MOS data from observations 1, 2 and 3 were affected by very high soft proton flaring. We only used events recorded during low background flaring for the spectral analysis and source identification. All events were used for timing and light curve analysis, and the task {\sc barycen} was run to correct all timing data to the Solar System barycentre. Figure \ref{lanxrayimg} shows the combined EPIC-pn and -MOS images from the four observations of Lanning 386. No results are presented using the RGS instrument due to the very low signal-to-noise ratio.

Photometry of Lanning 386 was also taken using the Galway Ultra Fast Imager (GUFI; \citealt{2011ASPC..448..219H}) mounted on the 1.8m Vatican Advanced Technology Telescope (VATT\footnote{The VATT telescope facility is operated by the Vatican Observatory, and is part of the Mount Graham International Observatory.}) on the nights of June 22 2009 (UT), December 4 2010 (UT) and December 6 2010 (UT). There were 1620 exposures taken on June 22 with a typical cadence of 10s, 471 exposures taken on Dec 4 with a typical cadence of 22s and 208 exposures on Dec 6 with a typical cadence of 42s. Reduction and analysis of these data was carried out using the typical {\sc Iraf}\footnote{{\sc Iraf} is distributed by the National Optical Astronomy Observatory, which is operated by the Association of Universities for Research in Astronomy (AURA) under cooperative agreement with the National Science Foundation} commands.

The Large Binocular Telescope (LBT) observed Lanning~386 with the Multi-Object Dual Spectrographs (MODS; \citealt{PoggeMODS}) on 2016 June 15 (UT) in binocular mode. This means MODS1 was taking data using the SX mirror (left side) simultaneously with MODS2 on the DX (right side), although the start times of the two spectrographs were not synchronised. MODS1 and MODS2 are nearly identical and in the dual grating mode both cover a wavelength range from 320 nm to 1$\micron$, divided into red and blue channels separated by a dichroic at 560 nm. The exposure for each spectrum was 120 s and the time between the start of consecutive exposures averaged 183 s. A slit width of 0.8 arcsec was used and the seeing averaged slightly better than 1.0 arcsec.

Thirty spectra were obtained from each spectrograph between 10:12 UT and 11:00 UT, providing a time resolution between 30 s and 100 s when the two sets of spectra are combined. The data reduction tasks were carried out using {\sc Iraf}.

\subsection{J1923}

The single \textit{XMM-Newton} observation of the field around J1923 failed to detect any source, with an upper limit on the flux from the region of $F_{X}<1.1\times10^{-14}$ erg cm$^{-2}$ s$^{-1}$ at the 2$\sigma$ level. Due to the non-detection of the source, no further results on J1923 are presented, except for a comparison to Lanning 386 in Section \ref{comp_both}.

\section{Timing Analysis}
\subsection{Optical Light Curve} \label{OM_Data}

\begin{figure}
	\includegraphics[width=80mm]{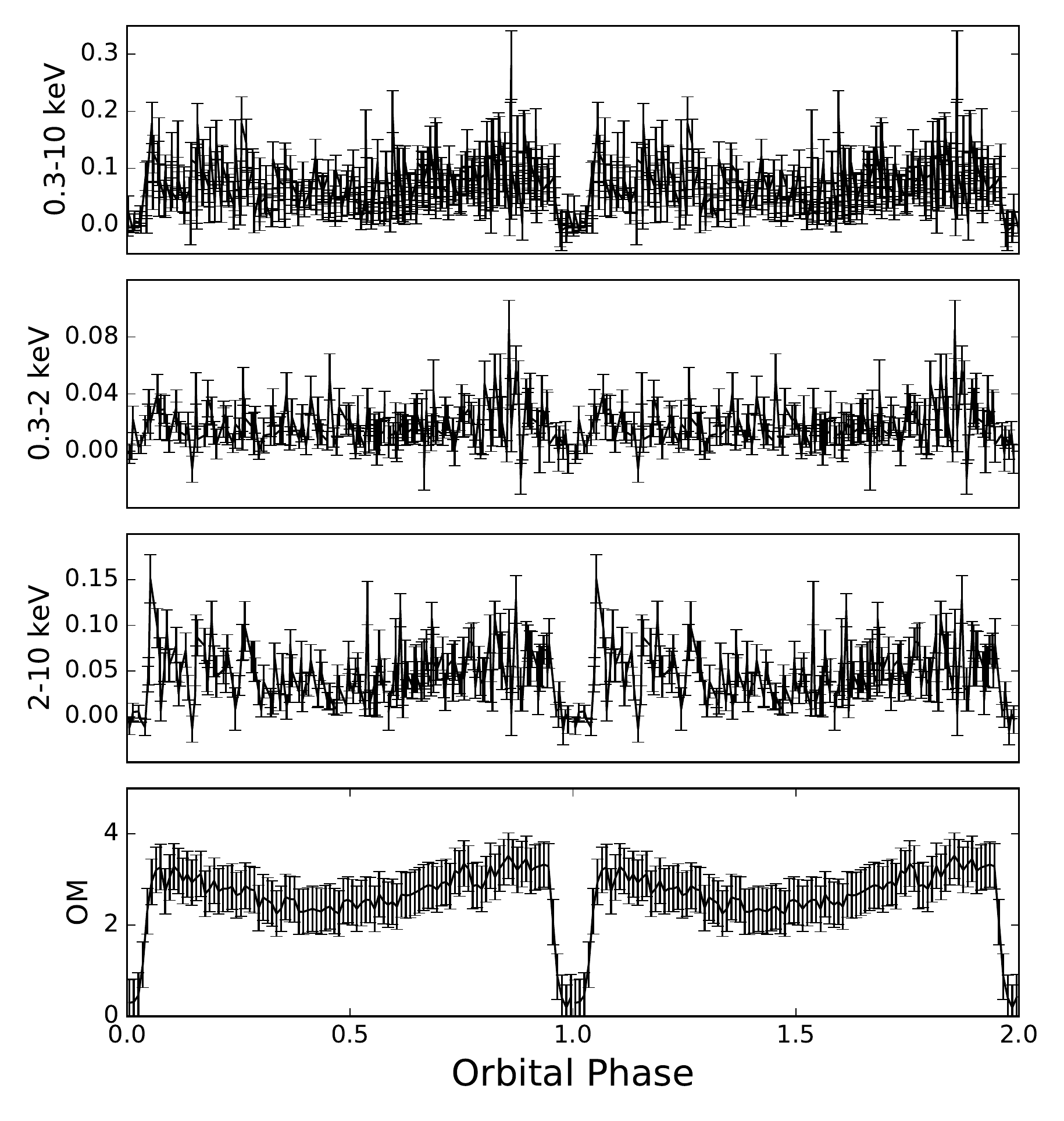}
	\caption{The extracted light curves of Lanning 386 from the EPIC-pn and OM data. The phase folded X-ray light curves for the full (0.3-10 keV, top panel), soft (0.3-2 keV, 2nd panel) and hard (2-10 keV, 3rd panel) spectral ranges are shown. The bottom panel is the phase folded optical data from the OM, taken in the V-band. The optical data has been binned with a bin width of 0.1 in phase. The y axis is in units of counts s$^{-1}$. Orbital phases have been repeated for clarity.}
	\label{lightcurves}
\end{figure}

The extracted optical light curve of Lanning 386 taken with the optical monitor can be seen in the bottom panel of Figure \ref{lightcurves}. The V-band magnitudes found in Table \ref{optical_phase_data} were calculated by taking the mean count rates of the individual observations and using the zero-point formula of

\begin{equation}
m = m_{zero}-2.5\log_{10}{\left(cts\right)}
\end{equation}

where $m_{zero}=17.9633$ for the V-band\footnote{XMM OM calibration document XMM-SOC-CAL-TN-0019}. The resulting magnitudes are consistent with the quiescent state of Lanning 386.

\begin{table*}
	\centering
	\caption{Properties of the optical observations of Lanning 386 from \textit{XMM-Newton} and the VATT.}
	\begin{tabular}{r c c c c c}
		\hline
		Observatory				&Obs No.		& Mean Count Rate	& Mag	& Mid-eclipse		& Cycle\\
								&				& cnts s$^{-1}$		&		& Time (BJD)		& No.\\
		\hline\hline
		VATT (V-band)			&2009 June 22   & -					& 16.76	& 2455004.937(1)	&8408\\
		VATT (V-band)			&2010 Dec 04  	& -					& 15.12	& 2455534.6598(5)	&11637\\
		VATT (U-band)			&2010 Dec 06	& - 				& 14.82 & 2455536.6285(5)	&11649\\
		\textit{XMM-Newton} OM	&1 				& 2.88				& 16.81	& 2457138.5955(5)	&21414\\
		\textit{XMM-Newton}	OM	&2				& 2.76				& 16.86	& 2457155.8218(5)	&21519\\
		\textit{XMM-Newton}	OM	&3				& 2.71				& 16.88	& 2457162.7110(5)	&21561\\
		\textit{XMM-Newton}	OM	&4				& 2.76				& 16.86	& 2457164.188(1)	&21570\\
		\hline
	\end{tabular}
	\label{optical_phase_data}
\end{table*}

Lanning 386 was undetectable during eclipse, as its eclipse magnitude of $\approx$18.8 mag  is below the flux sensitivity of the OM. To determine the mid-eclipse times, a Gaussian was fit to the ingress and egress of the eclipse visible in the optical observations. These mid-eclipse times were added to the mid-eclipse times calculated from the GUFI observations presented in Section \ref{ObsSect}, the GUFI observations presented in \cite{kennedy16} and the times given in Table 5 of \cite{Brady2008} to determine a new ephemeris. The best fit linear ephemeris was 

\begin{equation}
T (BJD) = 2453625.5876(1) + 0.16405193(2) E
\end{equation}

The O-C for these mid-eclipse timings is flat, with no detectable $\dot{P}$.

\subsection{X-ray Light Curves}
X-ray light curves of Lanning 386 and a nearby background region were extracted from the EPIC-pn data using the {\sc evselect} tool in SAS, and then the Lanning 386 light curves were corrected using the {\sc epiclccorr} tool. Light curves were extracted for the full (0.3-10 keV), soft (0.3-2 keV) and hard (2-10 keV) spectral ranges, and phased using the linear ephemeris found in Section \ref{OM_Data}. The resulting light curves can be seen in the top and middle panels of Figure \ref{lightcurves}. The X-ray light curves have a bin width of 200s. 

There is an eclipse visible at the expected phase in the full X-ray light curve, beginning at orbital phase 0.94 and lasting until orbital phase 0.05. The eclipse is the same width as the optical eclipse, suggesting that the source of the X-ray emission is eclipsed for the same duration as the dominant source of the optical emission in the system, which is thought to be the accretion disk. For the X-ray source of the system to be eclipsed, the inclination of this system must be high. The same eclipse is visible in the hard X-ray light curve (Figure \ref{lightcurves}, third panel), with the same eclipse width.

\subsection{Power Spectrum}
\begin{figure}
	\includegraphics[width=80mm]{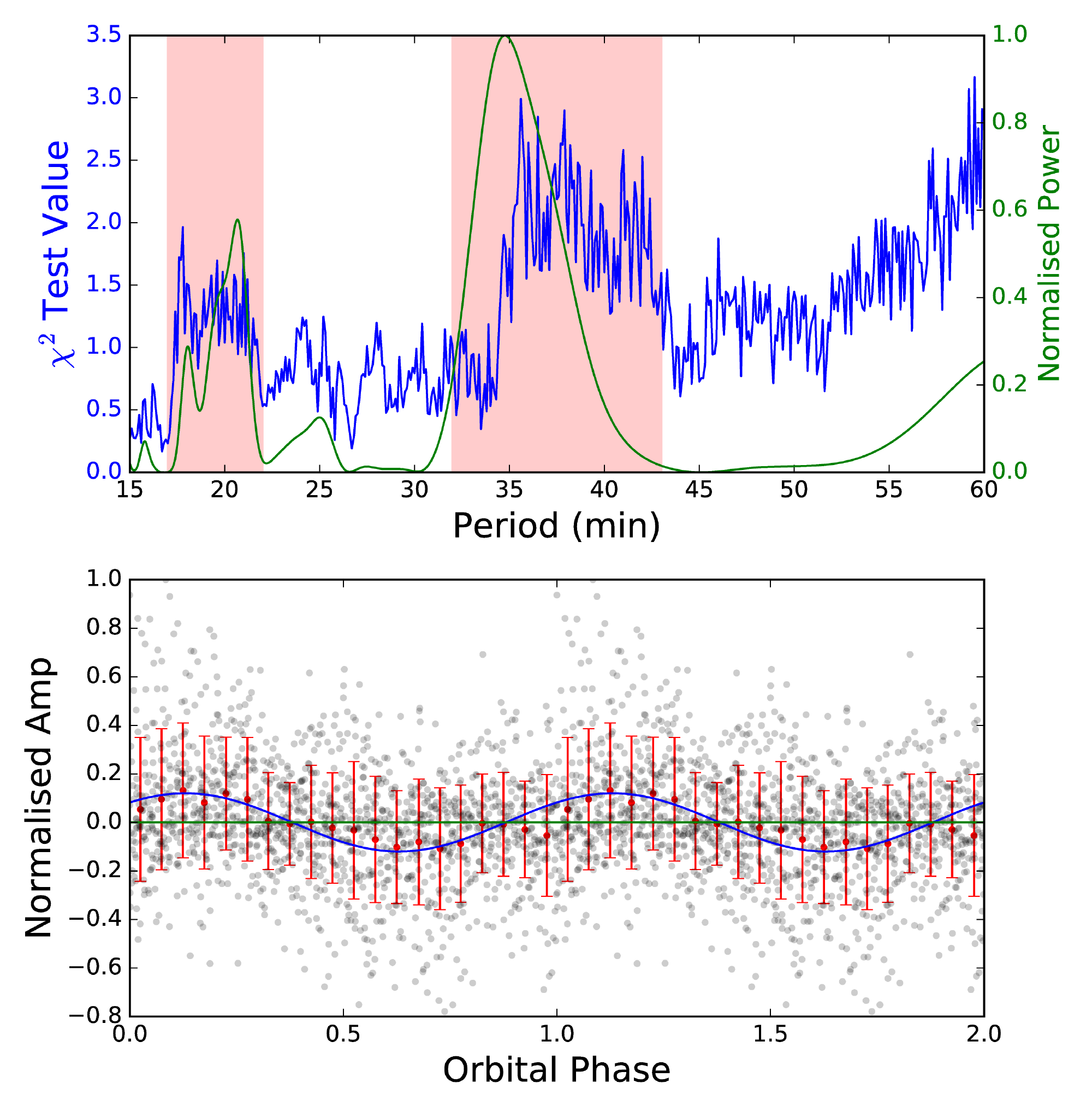}
	\caption{\textit{Top:} The $\chi^{2}$ test values (blue) and normalised Lomb Scargle power (green) of the Lanning 386 GUFI data. In both the $\chi^{2}$ test and power spectrum, there are strong periods detected between 17-22 min and 32-43 min (both of these regions are highlighted in red). \textit{Bottom:} The GUFI data phased on a period of 17.8 minutes (grey), and the data binned with a bin width of 0.05 in phase (red). The data shows a single cycle when phased on the 17.8 min period, and two cycles when phased on a 35 min period, suggetsing the shorter period is the period of the QPO. An example sine curve is plotted (blue) for comparison.}
	\label{gufi_powerspectra}
\end{figure}

Each night of the GUFI optical data was subjected to a Lomb-Scargle periodogram (LSP, \citealt{Lomb76}; \citealt{scargle82}) using the astroML python library \citep{astroML}. The three resulting power spectra were multiplied by each other, such that common peaks between the 3 power spectra would be enhanced, while noise peaks would be diminished. The resulting normalised power spectrum can be seen in the top panel of Figure \ref{gufi_powerspectra}. We also applied the $\chi^{2}$ period test to the GUFI data to test the validity of the periods, also shown in the top panel of Figure \ref{gufi_powerspectra}. The $\chi^{2}$ test works by binning the data, phasing the binned data to a range of periods and calculating the $\chi^{2}$ value for the binned data versus a normal distribution. When no period is present, small $\chi^{2}$ values are expected, while large $\chi^{2}$ values are expected when a periodic feature is present.  The strongest peaks in the power spectrum were between 17-22 min and 32-43 min. Phasing the data on different periods in these bands revealed a single cycle was present when the data was phased with a period in 17-22 min range, similar to that of J1923.

\begin{figure}
	\includegraphics[width=80mm]{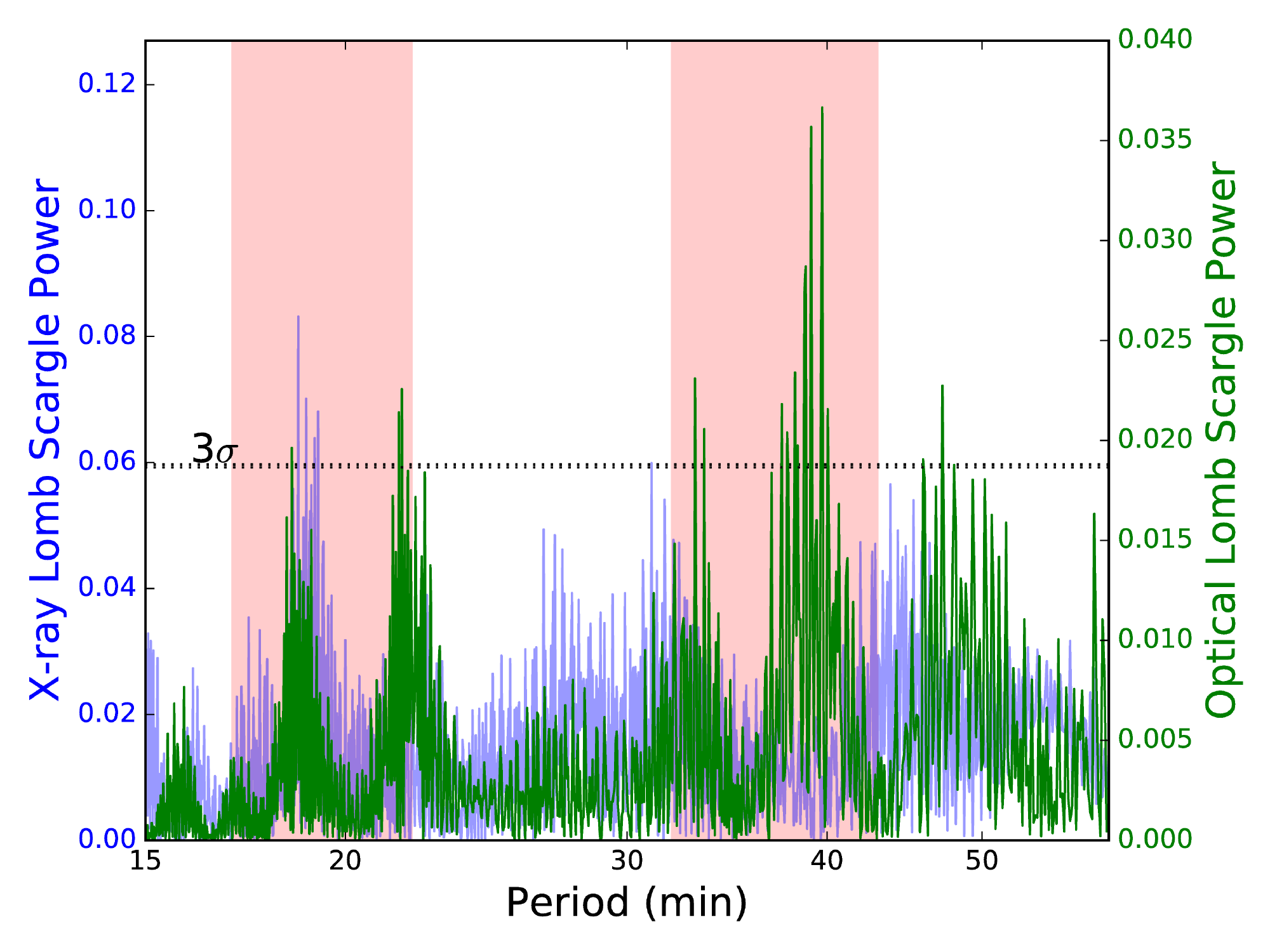}
	\caption{The optical (green) and X-ray (blue) power spectra of Lanning 386. There are two strong peaks at 18.7 min and 21.5 min in the optical spectrum which are detected the 3 $\sigma$ level. There is one peak that is also present in the X-ray power spectrum at 18.7 min. The red shaded areas correspond to the same shaded areas as in Figure \ref{gufi_powerspectra}.}
	\label{powerspectra}
\end{figure}

We also subjected the \textit{XMM-Newton} OM and 0.3-10 keV X-ray light curves to a LSP.  The resulting power spectra can be seen in Figure \ref{powerspectra}. The periodogram was bootstrap re-sampled \citep{Suveges2012} 2000 times to determine the 1$\sigma$, 2$\sigma$ and 3$\sigma$ levels.

The optical power spectra shows 2 broad peaks with periods centred on 18.5 and 22.5 minutes which are detected at the 3$\sigma$ level.

There is a sharp peak in the X-ray power spectrum at 18.7 minutes which was detected above the 3$\sigma$ level.This X-ray peak is as wide as the optical QPO peak at 18.5 minutes. Due to the position and shape of this peak, we consider it to be real and probably related to the optical QPOs.

The GUFI, OM and X-ray data all exhibit a similar QPO, with a periodicity of around 18.5 minutes. These peaks occur at the same periods as the QPOs seen in the similar object J1923, where the QPOs have a period of $\approx$ 20 minutes \citep{kennedy16}. Based on this, we attribute the periods in the optical and X-ray power spectrum to be a result of the QPOs present in this system.

\section{Spectral Analysis}
\subsection{Optical} \label{OptAnal}
\begin{figure}
	\includegraphics[width=80mm]{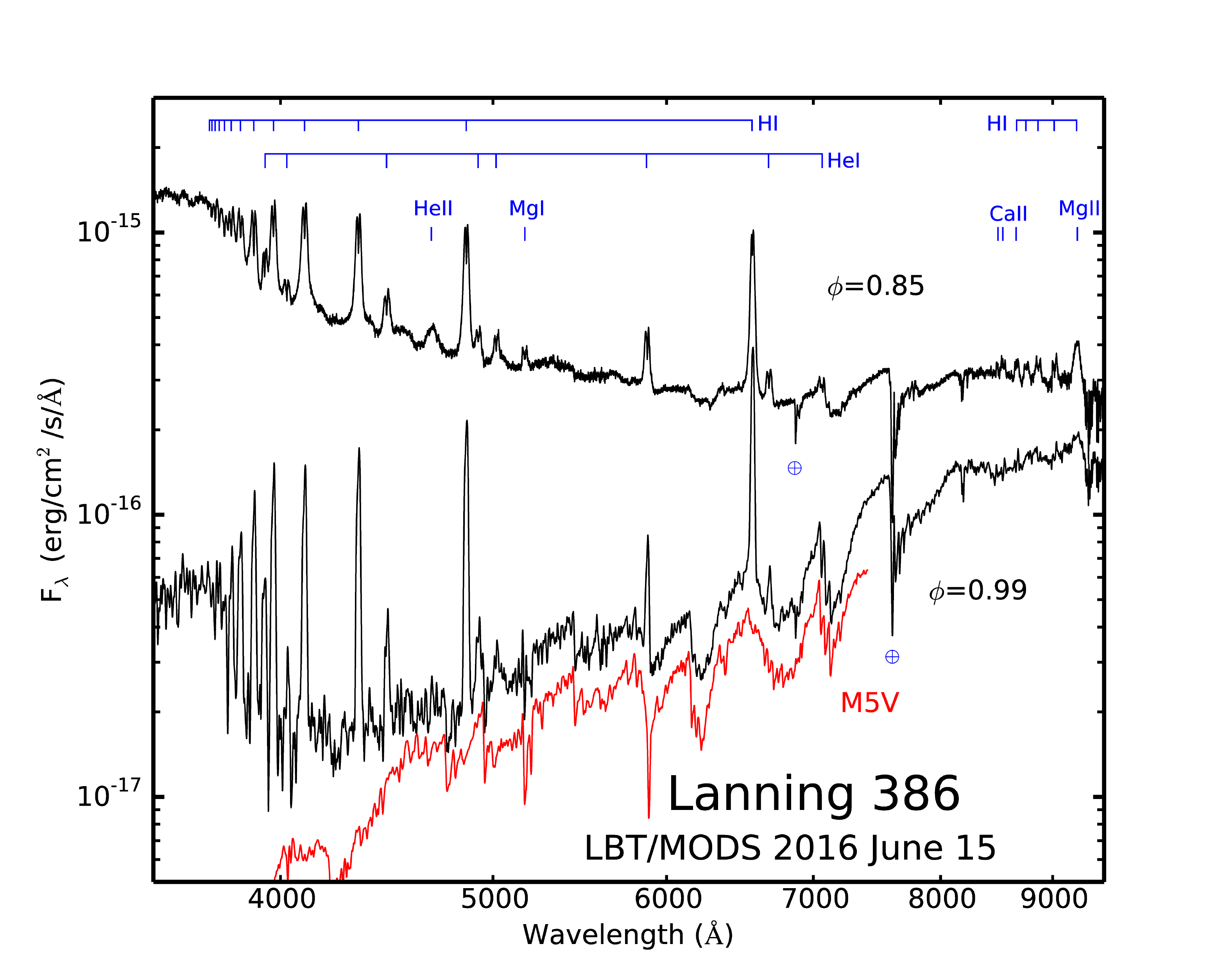}
	\caption{The spectrum of Lanning 386 taken at 2 different orbital phases using the MODS spectrographs on the 8.4 m LBT. The spectrum, taken during quiescence, shows strong \ion{H} and \ion{He}{I} emission lines. The lines are double peaked outside of eclipse, and there is a weak \ion{He}{II} component visible. The red line is a template M5V, which shows that most of the light in the continuum beyond 5000\AA is coming from the late-type secondary.}
	\label{opticalspectrum}
\end{figure}

\begin{figure*}
	\includegraphics[width=80mm]{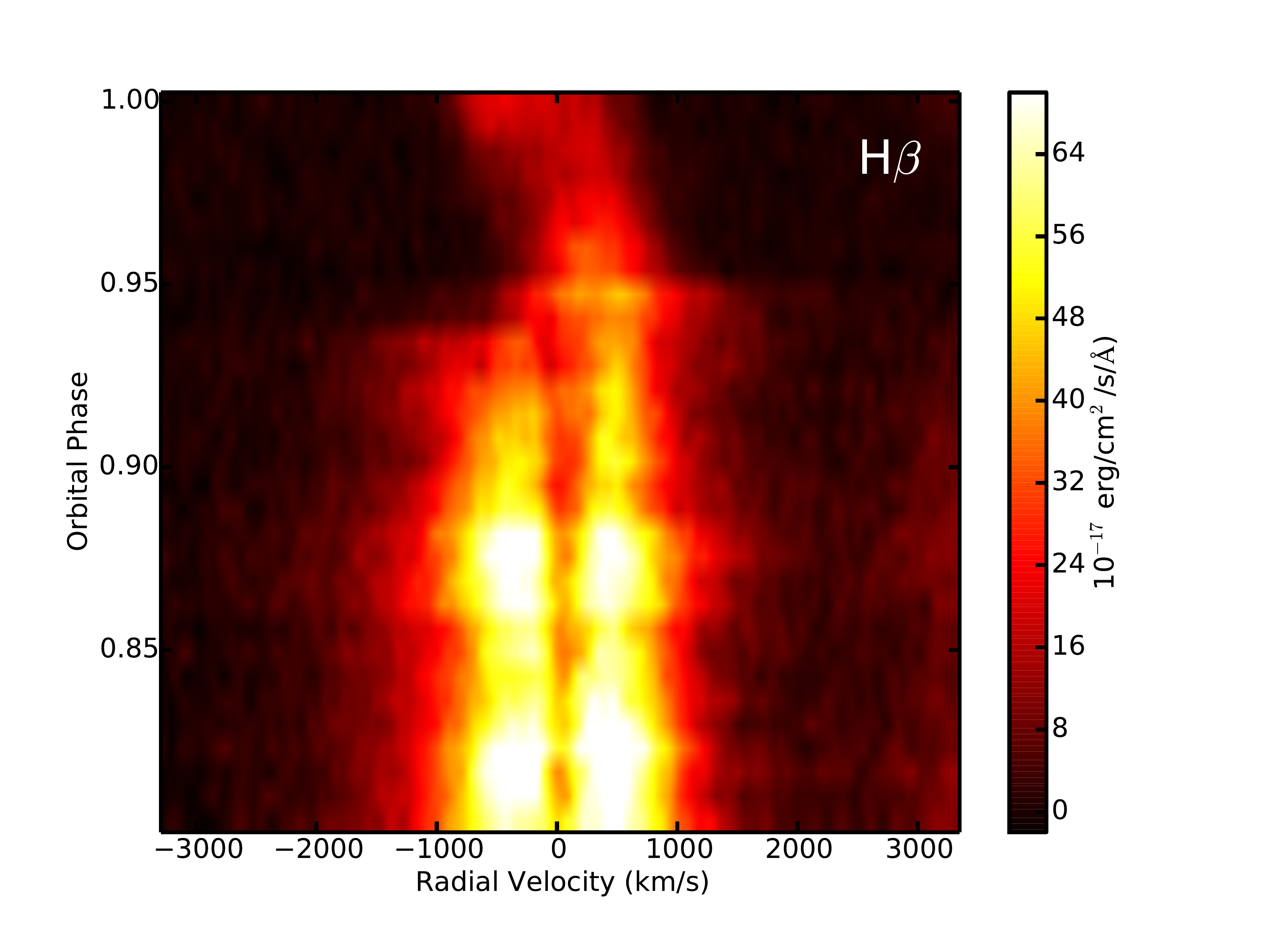}
	\includegraphics[width=80mm]{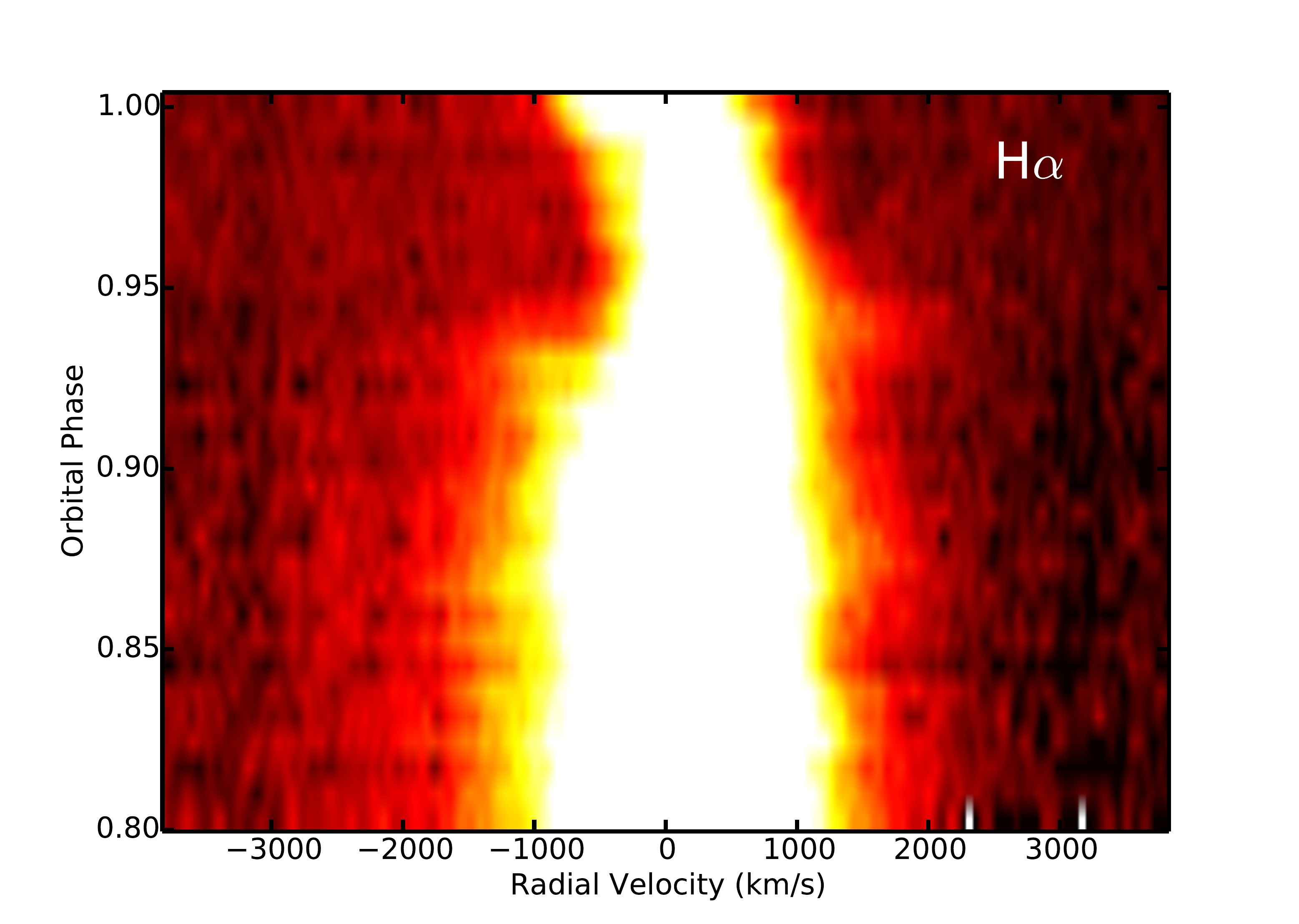}
	\caption{\textit{Left:} The trailed spectrum of H$\beta$ for Lanning 386 shows the lines are clearly double peaked outside of eclipse and during quiescence. As the system enters eclipse, the blue wing of the emission line is eclipsed first. \textit{Right:} The high contrast trailed spectrum of H$\alpha$ shows how quickly the blue wing of the H$\alpha$ line is eclipsed, with all blue emission eclipsed by phase 0.95. The red wing starts a shift towards 0 km s$^{-1}$ as the secondary begins to eclipse the red-shifted part of the accretion disk.}
	\label{trailedspectrum}
\end{figure*}

\begin{figure}
	\includegraphics[width=80mm]{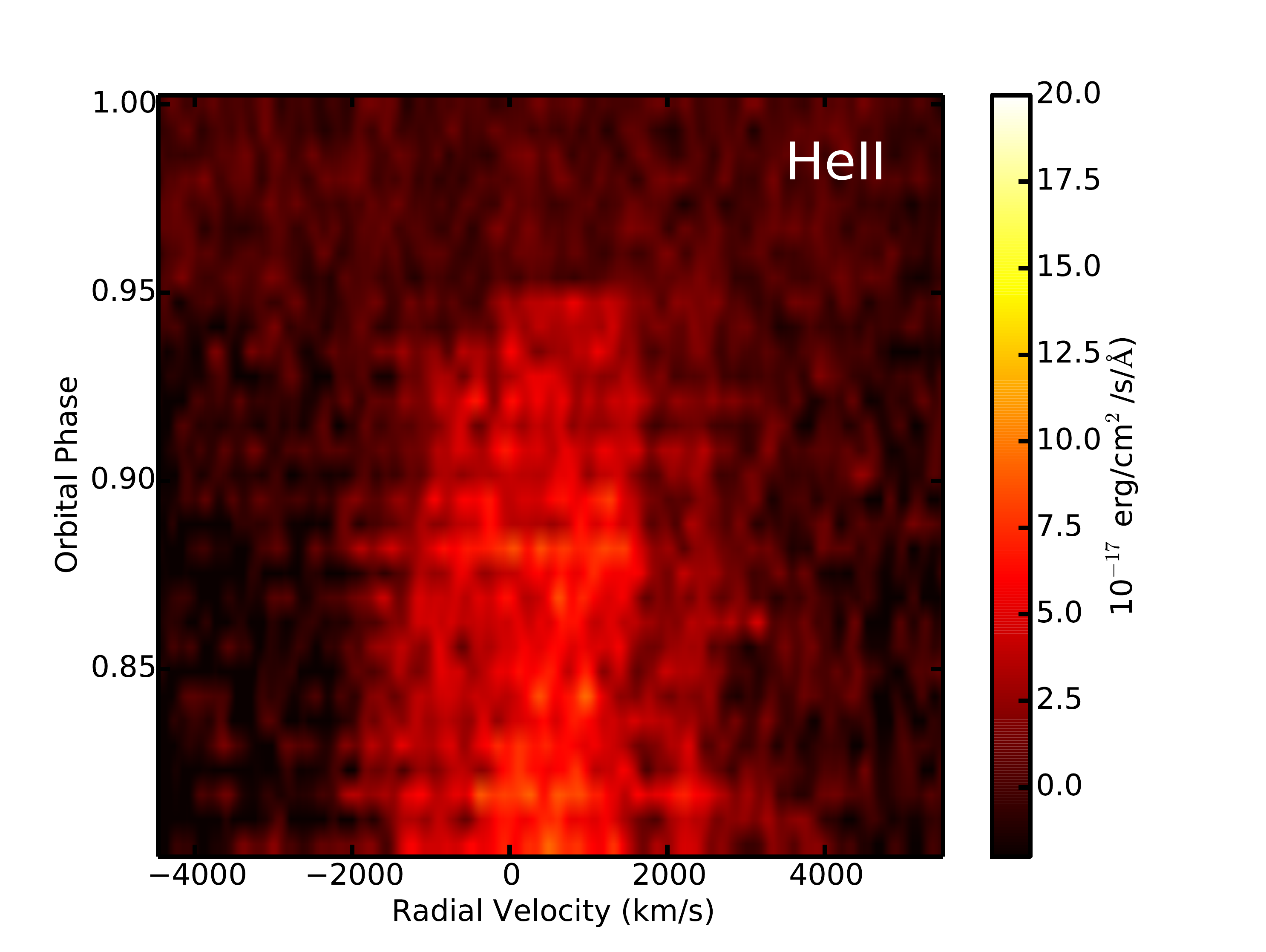}
	\caption{The trailed spectrum of \ion{He}{II$_{\lambda4686}$} shows that weak \ion{He}{II} emission is present in the system even in quiescence, and becomes fully eclipse around the same time as the H$\beta$ emission. As with the H$\alpha$ line, the \ion{He}{II} is eclipsed very quickly, only taking $\sim$0.01 phase for the line to become undetectable. There is an absorption feature present at 2000 km s$^{-1}$, which is consistent with an absorption feature from an M type star.}
	\label{trailedspectrum_He}
\end{figure}

The series of spectra obtained by the LBT begin at phase 0.80 and end at mid-eclipse (phase 1.00). Figure \ref{opticalspectrum} shows the average spectrum out of eclipse and the spectrum close to maximum eclipse. The ultraviolet and blue continua are seen to fade by a factor of 30 in brightness. The Balmer and HeI lines remain visible during eclipse and fade only by a factor of six in total flux. The spectral features of an M5-type secondary star become more prominent in eclipse. Using the MILES stellar library\footnote{\url{http://www.iac.es/proyecto/miles/}}, we find the M5V star HD173740 \citep{Cenarro2007} is a good fit to the secondary star spectrum and that the spectral type of the companion in Lanning 386 is likely in the M4-M5V range. This is slightly later than the M3.5e star suggested by \cite{Brady2008}.

The acquisition images and calibrated flux out of eclipse indicate that Lanning~386 was fainter than $V>17.3$ at the time of the LBT spectra and therefore in the quiescent state (as were the X-ray observations). Outside of eclipse, the Balmer and HeI lines are double peaked which can be due to the accretion disk or absorption by an accretion curtain if the system is magnetic. 

During outburst, Lanning~386 shows a very strong \ion{He}{II} plus Bowen blend emission feature but \ion{He}{II} is significantly weaker in quiescence according to \cite{Brady2008}. Our spectra clearly shows a weak but broad HeII line out of eclipse and no detection at mid-eclipse. The velocity width of \ion{He}{II} is 3400 km s$^{-1}$ (FWHM) compared with 1800 km s$^{-1}$ for the H$\beta$ emission. 

With 30 optical spectra taken over a phase range of 0.2, the typical temporal resolution is 2 minutes (0.008 in phase).  Figure \ref{trailedspectrum} shows the trailed spectra of the H$\alpha$ and H$\beta$ lines after the continuum has been subtracted. Some variability in the line flux is seen out of eclipse that may be due to a QPO seen in the X-ray and optical light curves. The double-peaked profile in H$\beta$ is clear and there is an obvious asymmetry between the red and blue shifted high-velocity tails. The blue emission is seen to extend beyond -2000 km s$^{-1}$ while the red side falls off more quickly. The scaling of H$\alpha$ has been set to clearly show the highest velocity emission.

As expected, the blue side of the emission line is the first to be eclipsed at phase 0.934, but its disappearance is particularly fast, blinking out in less than 0.01 in phase. A large fraction of the redshifted component is eclipsed at phase 0.948. A low-velocity redshifted component survives, transforming to a slightly blueshifted emission at mid-eclipse. The phase at which the blue side of the emission line is eclipsed also matches the phase at which the X-ray light curve beings to eclipse.

The fast disappearance of the blue side of the emission suggests that it comes from a very small region. The secondary motion of 0.05 radians means the size of the emission region is less than 6\% of the separation between the two stars. The time between the blue and red sides being eclipsed means they are separated by 9\% of the orbital radius. After the eclipse there remains a weak Balmer emission moving from red to blue, but it is not clear from where this light is originating.

Figure \ref{trailedspectrum_He} shows the trailed spectrum for \ion{He}{II}. It shows a similar rapid eclipse of the blue side (phase 0.936) and then red side (phase 0.948). There are no further emission components beyond phase 0.948 and we conclude the eclipse is total for the \ion{He}{II} line.

\subsection{X-ray}\label{SpecAnal}
\begin{figure}
	\includegraphics[width=80mm]{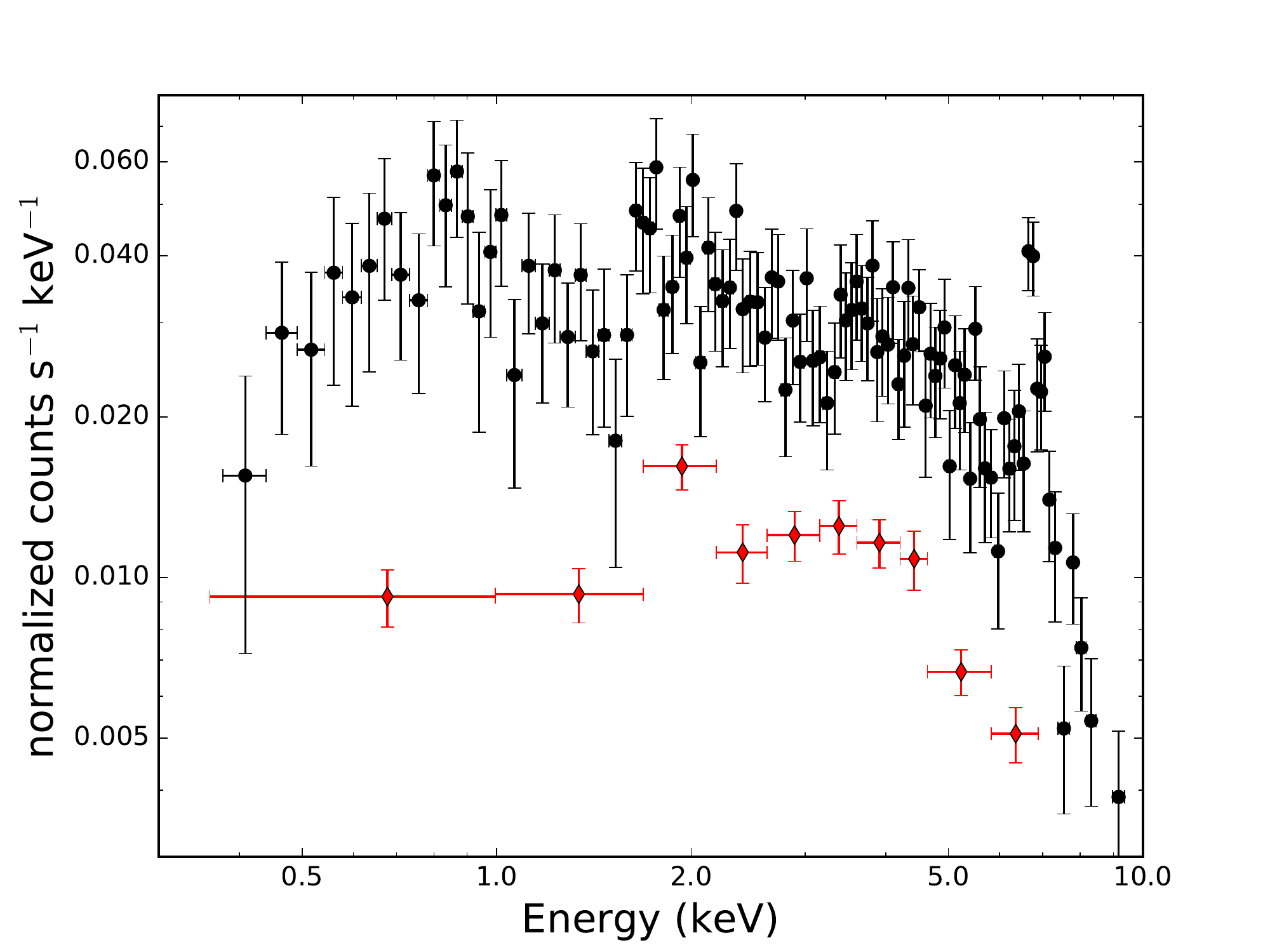}
	\caption{The average X-ray spectrum of Lanning 386. The black circles are the EPIC-pn spectrum, while the red diamonds are the EPIC-MOS data. There are 2 blended strong emission lines in the EPIC-pn spectrum at 6.7 keV and 6.9 keV, which match the energies of \ion{Fe} {xxv} and \ion{Fe}{xxvi}.}
	\label{rawsxrayspectrum}
\end{figure}

X-ray spectra were extracted for the energy range 0.3-10 keV for all 4 observations of Lanning 386 using the EPIC-pn and EPIC-MOS instruments. The average X-ray spectrum from both instruments is shown in Figure \ref{rawsxrayspectrum}. The spectra have been rebinned to have 25 counts per spectral bin.

The data were analysed using the HEASARC software package {\sc Xspec} (version 12.9.0; \citealt{1996ASPC..101...17A}). Analysis was limited to the EPIC-pn spectrum, as the -MOS spectrum had a low signal-to-noise ratio.

Initially, the data were fit using an absorbed power law model ({\sc tbabs*powerlaw}). The resulting fit showed a hard spectrum with a spectral index $\Gamma=0.32\pm0.05$ and a column density of $N_{H}=0.05\pm0.03 \times 10^{22}$ cm$^{-2}$, but the fit was poor, with a $\chi^{2}_{R}=2.00$. We next added 2 {\sc Gaussian} components to try and fit the 6.701 keV  \ion{Fe}{xxv} and 6.97 keV \ion{Fe}{xxvi} line. The resulting fit had a $\chi^{2}_{R}=1.66$ and is listed as Model 1 in Table \ref{model_data}.

The most common model used to describe the X-ray emission of a cataclysmic variable is a single temperature thin thermal plasma ({\sc Mekal} in {\sc Xspec}). Fitting an absorbed {\sc Mekal} model to the data gave a $\chi^{2}_{R}=2.75$ and a high $N_{H}=1.7\pm0.1 \times 10^{22}$ cm$^{-2}$. However, there were large, systematic residuals below 2 keV for this fit. This suggests a single temperature model with simple absorption is insufficient to model this source. We next attempted to fit a multi temperature model by adding a second {\sc Mekal} component to fit the spectrum below 2 keV. The result was slightly poorer than the simpler model, with a $\chi^{2}_{R}=2.81$ and a similar column density of $N_{H}=1.7\pm0.1 \times 10^{22}$ cm$^{-2}$.

\begin{figure}
	\includegraphics[width=80mm]{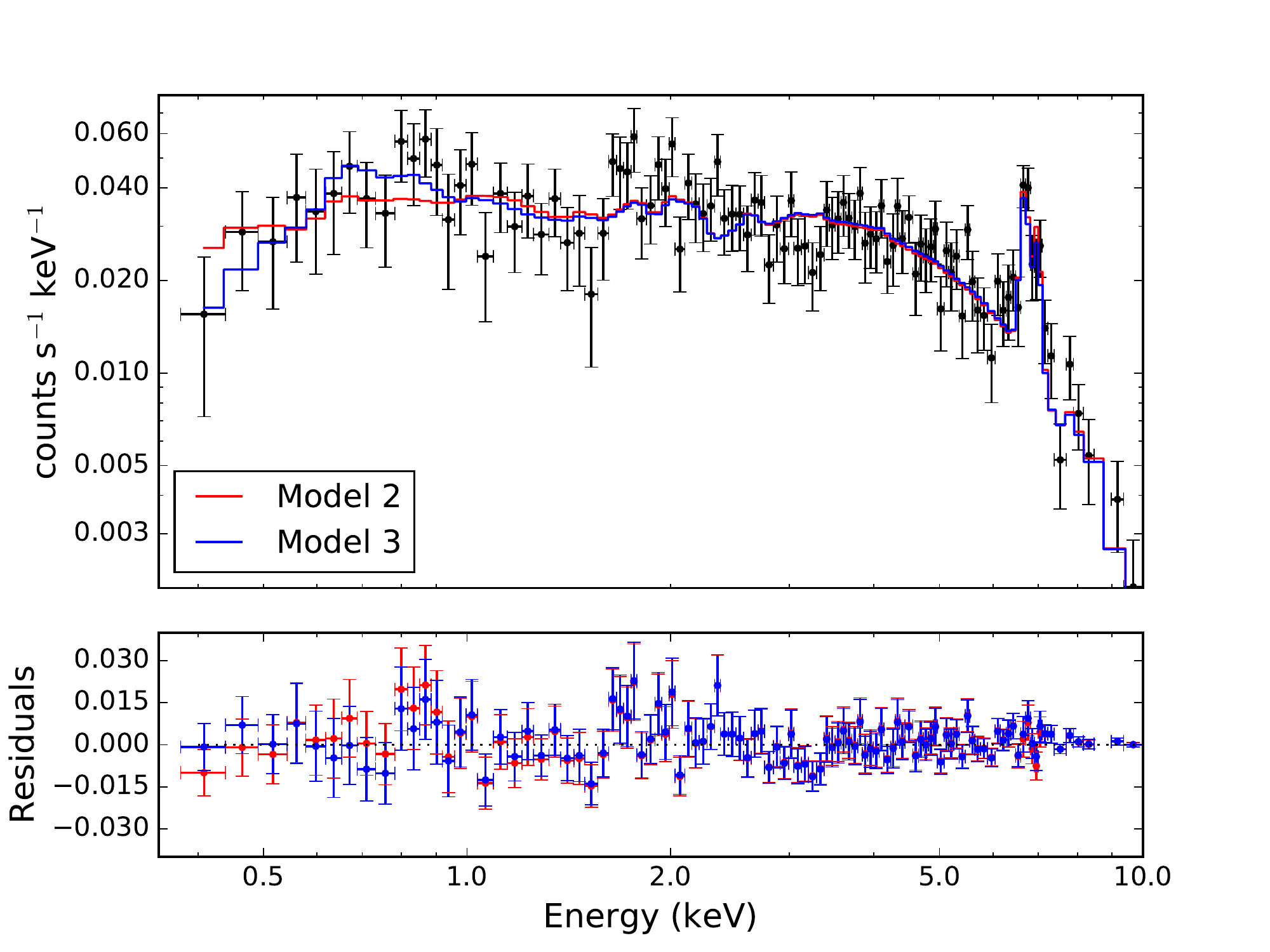}
	\caption{The 2 best fit models to the EPIC-pn spectrum. Model 2 (red) is a single temperature model with a lower than expected interstellar column density. Model 3 (blue) has a column density fixed at the interstellar value found by \cite{Eracleous2002}, but requires two different plasma temperatures (a hot and cold component) to fit the spectrum. The details of both model can be seen in Table \ref{model_data}.}
	\label{fitxrayspectrum}
\end{figure}

IPs are often modelled using a single {\sc Mekal} component along with a complex absorption model \citep{Staude2008}. Following this approach, we next fit Lanning 386 using the relatively complex model of {\sc Tbabs}*({\sc PartCov*Tbabs})*({\sc Mekal}). Here, the first {\sc Tbabs} component accounts for interstellar absorption, ({\sc PartCov*Tbabs}) models the circumstellar absorption and {\sc Mekal} is as before. As in \cite{Nasiroglu2012}, our first fit could not constrain the plasma temperature, but provided a lower limit of $>8$ keV at the 3$\sigma$ level. The resulting fit did not model the 6.7 keV \ion{Fe}{xxv} line well, and had a $\chi^{2}_{R}=1.04$. Finally, we allowed the metal abundance of the {\sc Mekal} component to vary, such that the model fit around the 6.7 keV \ion{Fe}{xxv} line might be improved. The final fit had a $\chi^{2}_{R}=0.84$ and can be seen in Figure \ref{fitxrayspectrum}. The best fit parameters are listed under Model 2 in Table \ref{model_data}.

\cite{Eracleous2002} used the strength of the $\lambda$4430 DIB to estimate an $A_{V}=1.5\pm0.2$ in the direction of Lanning 386. Using the linear relation between column density and $A_{V}$ given by \cite{Guver2009}, this equates to a column density of $N_{H}=3.3\pm0.4\times 10^{21}$ cm$^{-2}$. This is higher than the column density of $N_{H}=2.54 \times 10^{21}$ cm$^{-2}$ from the Leiden/Argentine/Bonn (LAB) Survey of Galactic HI \citep{2005A&A...440..775K}\footnote{Calculated using the online $N_{H}$ tool available at \url{https://heasarc.gsfc.nasa.gov/cgi-bin/Tools/w3nh/w3nh.pl}}.

The column density of the interstellar absorber in Model 2 ( $N_{H}=(0.8^{+1.0}_{-0.7})\times 10^{21}$ cm$^{-2}$ ) is far below the values from \cite{Eracleous2002} and the LAB survey. Motivated by this descrepancy, the column density was fixed to 3.3$\times 10^{21}$ cm$^{-2}$, and the X-ray spectrum refit using Model 2. This model had large errors in the soft X-rays, and a $\chi^{2}_{R}=1.2$. As such, we added a second {\sc Mekal} component with a low plasma temperature. The resulting model, {\sc Tbabs}*({\sc PartCov*Tbabs})*({\sc Mekal}+{\sc Mekal}) is listed under Model 3 in Table \ref{model_data}, and shown in Figure \ref{fitxrayspectrum}. The best fit had a $\chi^{2}_{R}=0.82$, and did not require a variable metallicity. In the following Sections, the {\sc Mekal} model with the lower plasma temperature of 0.24$^{+0.17}_{-0.08}$ keV is referred to as the ``cool'' plasma component, and the {\sc Mekal} model with the higher plasma temperature of 9$^{+4}_{-2}$ is the ``hot'' plasma component

\begin{table*}
	\centering
	\caption{Parameters obtained for various models in {\sc Xspec} using the Lanning 386 data. \textit{(a) parameter frozen at this value for the fit}}
	\begin{tabular}{r c c c c}
		\hline
		Component 						& Parameter							& Model 1 	& Model 2					& Model 3 \\
		\hline\hline
		{\sc tbabs (Interstellar)}		& $N_{H}$ (10$^{22}$ cm$^{-2}$)		& 0.06(3)	& 0.08$^{+0.10}_{-0.07}$	& 0.33$^{a}$\\
		\\
		{\sc tbabs (Stellar)}			& $N_{H}$ (10$^{22}$ cm$^{-2}$)		& -			& 4.0$^{+1.7}_{-1.1}$ 		& 4.8$^{+2.0}_{-1.4}$\\
		{\sc partcov}					& cvf								& - 		& 0.86$^{+0.05}_{-0.07}$	& 0.88$^{+0.04}_{-0.06}$\\
		\\
		\multirow{2}{*}{\sc Powerlaw}	& $\Gamma$							& 0.37(6)	& -							& -\\
										& norm ($\times 10^{-5}$)			& 1.13(9)	& -							& -\\		
		\\					
		\multirow{3}{*}{\sc Mekal}		& kT (keV)							& -			& 9$^{+5}_{-2}$				& 9$^{+4}_{-2}$\\\\		
										& Abundance							& -			& 1.2$^{+0.96}_{-0.60}$		& 1.0$^{a}$	\\
										& norm ($\times 10^{-3}$)			& -			& 0.35(2)					& 0.35(2)\\
		\\
		\multirow{3}{*}{\sc Mekal}		& kT (keV)							& -			& -							& 0.24$^{+0.17}_{-0.08}$\\\\		
										& Abundance							& -			& -							& 1.0$^{a}$	\\
										& norm ($\times 10^{-3}$)			& -			& -							& 0.2(1)\\
		\\
		\multirow{2}{*}{\sc Gaussian1}	& Line Centre (keV)					& 6.70$^{a}$& -				            & -\\
										& Sigma (keV)						& 0.1(1)	& -					        & -\\
		\\
		\multirow{2}{*}{\sc Gaussian2}	& Line Centre (keV)					&6.97$^{a}$ & -							& -\\
										& Sigma (keV)						&0.1(1)		& -							& -\\
		\\
										& $\chi^{2}_{R}$					& 1.66		& 0.84						& 0.82\\
		\hline
	\end{tabular}
	\label{model_data}
\end{table*}

The unabsorbed flux in the 2-10 keV X-ray spectral range, F$_{X(2-10)}$, was found to be $\left(5.7\pm0.8\right)\times$10$^{-13}$ erg cm$^{-2}$ s$^{-1}$ by multiplying the {\sc Mekal} components by the {\sc Cflux} model in {\sc Xspec}, which estimates the flux of a given component over a given energy range. 

\subsubsection{Equivalent Widths and intensities of the Fe lines}
We measured the equivalent widths of the 6.7 keV \ion{Fe}{xxv} and 6.97 keV \ion{Fe}{xxvi} lines using the {\sc Eqwidth} command in {\sc Xspec} after fitting a {\sc Gaussian} model to both of the lines. The equivalent width of the 6.7 keV line was 420$\pm$50 eV, and the equivalent width of the 6.97 keV line was 200$\pm$50 eV. The intensity of both of these lines was also estimated by wrapping the {\sc Gaussian} models in the {\sc Cflux} model in {\sc Xspec}. The intensities were measured to be I$_{6.7}=2.4\pm0.7 \times 10^{-14}$ erg cm$^{-2}$ s$^{-1}$ and I$_{6.97}=1.3\pm0.5 \times 10^{-14}$ erg cm$^{-2}$ s$^{-1}$.

To determine an upper limit on the flux from the 6.4 keV \ion{Fe} line, a {\sc Gaussian} and {\sc Powerlaw} model was fit to the region around 6.4 keV. The resulting upper limit on the 6.4 keV line was I$_{6.4}<3.2\times 10^{-15}$ erg cm$^{-2}$ at the 3$\sigma$ level. The upper limit on the equivalent width of the 6.4 keV \ion{Fe} line was $<50$ eV. 

\subsection{An updated distance measurement}
\cite{Brady2008} estimated the distance to Lanning 386 to be $\sim$ 500 pc using a spectral type of M3.3 for the secondary and assuming a reddening of $E(B-V)<0.2$ ($A_{V}=0.62$ mag). The new best matching spectral type  of M5V gave an absolute magnitude for the companion in the system of $M_{V}=13.2\pm0.3$, using the relation between spectral type and surface brightness from \cite{Beuermann2006}. Combined with a visual extinction of $A_{V}=1.5$ \citep{Eracleous2002}, the distance to Lanning 386 was determined to be 160$\pm$50 pc.

Using this distance we derive an (unabsorbed) X-ray luminosity of $L_{X(2-10)}=\left(1.6^{+2.0}_{-0.9}\right)\times$10$^{30}$ erg s$^{-1}$.

\section{Discussion}
\subsection{Is Lanning 386 magnetic?} \label{IsItMagnetic}
Lanning 386 is thought to be an SW Sextantis type star, which are suspected to be magnetic CVs with very high accretion rates. The light curve of Lanning 386, which shows frequent outbursts and QPOs, suggests such a high accretion rate.

From the X-ray modeling of Lanning 386, we favour Model 3 over Model 2, as this model, by design, had an interstellar absorber equal to that measured by \cite{Eracleous2002}. This model also had the added bonus of not requiring a variable metallicity. In the following, we limit our discussion to Model 3.

The X-ray spectrum of Lanning 386 was very hard, and is typical of IPs, but also of nova-like CVs, which are non-magnetic CVs undergoing high mass transfer rates \citep{Balman2014}. The hot plasma temperature of $kT=9^{+4}_{-2}$ keV is not atypical of the plasma temperature found in IPs, arising from the interaction between the accretion curtain and the white dwarf \citep{2006csxs.book..421K}. However, we do note that this plasma temperature is lower than the majority of IPs. The soft {\sc Mekal} component, with a plasma temperature of $kT=0.24^{+0.17}_{-0.08}$ keV, has also been seen in other IPs. Additionally, the requirement of a partial covering absorption component is common for an IP (\citealt{Evans2007}; \citealt{Bernardini2012}; \citealt{Bernardini2015}). Table \ref{IP_data_list} shows the partially covered, absorbed Bremsstrahlung/mekal fits to a range of IPs from the literature. We note that Lanning 386 and FO Aqr have the highest partial covering fraction of all of the IPs listed in Table \ref{IP_data_list}.

\begin{table*}
	\centering
	\caption{Model fits to different IPs, showing the partial covering fraction, circumstellar column density and plasma temperature}
	\begin{tabular}{l c c c c r}
		\hline
		Name						&  \multicolumn{2}{c}{Partial Covering} 									& $kT_{cold}$ 			& $kT_{max}$						& Reference\\
									& n$_{\textnormal{H}}$ ($10^{23}$cm$^{-2}$)		& CvrFract					& keV 					& keV								& \\
		\hline\hline
		RXS J070407.9+262501		& 1.2$^{+1.2}_{-0.6}$							& 0.54$^{+0.07}_{-0.09}$	& 0.088$\pm0.005$		& 11$^{+13}_{-3}$				& \cite{Anzolin2008}\\
		RXS J180340.0+401214		& 5$^{+5}_{-2}$									& 0.5$^{+0.3}_{-0.2}$		& 0.10$\pm0.02$			& 12$^{+27}_{-3}$				& \cite{Anzolin2008}\\
		MU Camelopardalis			& 0.79$\pm0.06$									& 0.61$\pm0.02$				& 0.059$\pm0.07$		& 35$\pm10$						& \cite{Staude2008}\\
		IGRJ04571+4527				& 0.54$\pm0.01$									& 0.32$\pm0.03$				& 6.8$^{+1.7}_{-1.2}$	& 63$^{+8}_{-19}$				& \cite{Bernardini2015}\\
		Swift J0525.6+2416 3.2.1	& 0.18$\pm0.02$									& 0.45$\pm0.02$				& 0.19$\pm0.01$			& 40$^{+9}_{-5}$				& \cite{Bernardini2015}\\
		IGR J08390-4833				& 0.68$\pm0.08$									& 0.60$\pm0.02$				& 0.69$\pm0.05$				& 54$\pm9$					& \cite{Bernardini2012}\\
		IGR J18308-1232				& 1.5$\pm0.2$									& 0.60$\pm0.03$				& 0.16$\pm0.01$				& 40$\pm20$					& \cite{Bernardini2012}\\
		IGR J16500-3307				& 1.1$\pm0.1$									& 0.62$\pm0.01$				& 0.18$\pm0.01$				& 59$\pm8$					& \cite{Bernardini2012}\\
		IGR J17195-4100				& 0.63$\pm0.03$									& 0.43$\pm0.01$				& 0.159$\pm0.005$			& 30$\pm2$					& \cite{Bernardini2012}\\
		V2069 Cyg					& 0.8$\pm0.1$									& 0.64$\pm0.02$				& 0.070$\pm0.004$			& 32$\pm4$					& \cite{Bernardini2012}\\
		RX J0636+3535				& 1.1$\pm0.1$									& 0.57$\pm0.03$				& 0.079$\pm0.002$			& 36$\pm5$					& \cite{Bernardini2012}\\
		IGR J15094-6649				& 0.66$\pm0.06$									& 0.54$\pm0.01$				& 0.12$\pm0.01$				& 34$\pm3$					& \cite{Bernardini2012}\\
		RXS J073237.6−133113		& -												& -							& -							& 37$\pm7$					& \cite{Butters2007}\\
		V405 Aur					& 0.61$^{+0.06}_{-0.05}$						& 0.54$\pm0.03$				& 0.168$^{+0.008}_{-0.003}$	& 9$\pm1$					& \cite{Evans2004a}\\
		UU Columbae					& 0.83$^{+0.52}_{-0.32}$						& 0.43$^{+0.15}_{-0.17}$	& 0.18$\pm0.01$				& 27$^{+\infty}_{-17}$		& \cite{Martino2006}\\
		FO Aqr						& 1.9$^{+0.12}_{-0.09}$							& 0.86$^{+0.02}_{-0.03}$	& 0.176$\pm0.002$			& 14$^{+7}_{-3}$			& \cite{Evans2004}\\
		WX Pyx						& -												& -							& -							& 18$^{+24}_{-6}$			& \cite{Schlegel2005}\\
		EX Hya						& 0.001-0.086									& 0.22-0.35					& 0.66$\pm0.01$				& 30-60						& \cite{Pekon2010}\\
		\hline
		Lanning 386					& 0.48$^{+0.20}_{-0.14}$						& 0.88$^{+0.04}_{-0.06}$	& 0.24$^{+0.17}_{-0.08}$	& 9$^{+4}_{-2}$				&\\
		\hline
	\end{tabular}
	\label{IP_data_list}
\end{table*}

There was a weak signal detected between 17-22 min in in the X-ray light curve of Lanning 386. This period was also detected in the optical monitor data and the archival GUFI data. \cite{Patterson2002} and \cite{Warner2004} have proposed that the cause of QPOs is the spin period of the WD but instead of obtaining a clean signal at a single period, the spin period is smeared out into a QPO by reprocessing of the rotating beam by a varying period QPO source. Hence, in the case of Lanning 386, this weak optical QPO may be a manifestation of the spin period of the WD. The low significance detection of the signal in the X-ray data could be due to the systems inferred high inclination, which might be blocking most of the spin signal from the WD. For IPs, it is common for the spin period to be detected in the X-rays and for the beat period 

\begin{equation}
P_{beat}=\frac{1}{\frac{1}{P_{spin}}-\frac{1}{P_{orb}}}
\end{equation}

to be detected in the optical. If we take the spin period of the WD to be 17.8 min, the beat period would be 19.25 mins, which is within the period range detected in the GUFI and OM data. This suggests that the spin period and beat period are indistinguishable in our data. If the 17.8 min signal is the spin period of the white dwarf, then $P_{spin}=0.08\: P_{orb}$, which is close to the $\frac{P_{spin}}{P_{orb}}=0.1$ value around which IPs seem to cluster \citep{Scaringi2010a}.

IPs typically have an observed $L_{X(2-10)}$ in the range 10$^{32}$-10$^{33}$ erg s$^{-1}$ (e.g. \citealt{Xu2016}). This is two orders of magnitude higher than our calculated luminosity of $L_{X(2-10)}=\left(1.6^{+2.0}_{-0.9}\right)\times$10$^{30}$ erg s$^{-1}$, suggesting Lanning 386 is very under-luminous in the 2-10 keV X-ray range.

A smaller discrepancy has been seen in IPs previously, and can be seen in DQ Her itself \citep{Patterson1994} and also in EX Hya, which is an IP \citep{Vogt1980} that displays some SX Sextantis properties (\citealt{Hellier2000}; \citealt{hsf03} \footnote{See D.~W. Hoard's Big List of SW Sextantis Stars at \url{http://www.dwhoard.com/biglist} \citep{hsf03}.}) and has an estimated $L_{X(2-10)}=4.13\pm0.13\times10^{31}$ erg s$^{-1}$. However, Lanning 386 is still between 10-15 times fainter than this. There are several factors which could be affecting the X-ray luminosity. The first, and most important, is the high inclination of this system, which may be leading us to underestiamte the X-ray luminosity of the source. It also could be due to Lanning 386 having a much lower magnetic moment in comparison to other IPs. We will return to this point in Section \ref{MagMom}.

If Lanning 386 is magnetic, then a lower bound on the white dwarf mass can be made using the higher plasma temperature from the {\sc Mekal} model in Model 3. The {\sc Mekal} component of our model can be replaced with the {\sc Ipolar} model of \cite{Suleimanov2016}, where the only input parameters are the white dwarf mass and the magnetospheric radius of the white dwarf. (Formerly, this was the {\sc Ipm} model of \cite{Suleimanov2005}, but was recently updated to include the magnetospheric boundary in the system). This model relies on the assumption that the maximum shock temperature in the accretion curtain of an IP is related solely to the white dwarf mass and magnetospheric boundary. Replacing the {\sc Mekal} component with {\sc Ipolar} gives a white dwarf mass $>$ 0.5 M$_{\odot}$, and does not constrain the magnetospheric boundary.

\subsubsection{Fe line comparisons}

Next, we compare our X-ray emission-line measurements to those in the recent compilation of \cite{Xu2016}. The low upper limit on the 6.4 keV equivalent width (EW), combined with the 6.7 keV/6.97 keV EW ratio, suggest a quiescent dwarf nova (DN) classification, consistent with the relatively low $L_{X}$ we measure for Lanning 386. On the other hand, the 6.97 keV/6.7 keV line intensity ratio of 0.6$\pm$0.3 for Lanning 386 is comparable to values measured for both DNs as a class and IPs as a class \citep{Xu2016}.  Indeed, the most comparable source in the list of measurements of \cite{Xu2016} is EX Hya (closest in EWs, intensity ratios and Lx), the previously mentioned IP with a possible SW Sex association.

Hence, we conclude that whilst the X-ray emission-line and continuum characteristics of Lanning 386 are difficult to reconcile with the average values of any particular CV sub-type, they are most consistent with those of the IP EX Hya.

\subsubsection{An upper limit of the magnetic moment of Lanning 386} \label{MagMom}

Based on the separation of blue and red emitting regions of the accretion disk of approximately 9\% of the orbital separation, the inner radius of the accretion disk is approximately 0.05$a$, where $a$ is the orbital separation. \cite{Patterson2002} estimated that, for an upper limit on inner disk radius of $R_{d}\leq 0.3a$, the magnetic moment is related to the period of the system and the mass accretion rate by
\begin{equation}
\label{patterson_mag_mom}
\mu_{33} \leq 0.2 \dot{M}_{16}^{1/2}P_{hr}^{7/6}
\end{equation}
where $\mu_{33}$ is the magnetic moment in units of 10$^{33}$ gauss cm$^{3}$, $\dot{M_{16}^{1/2}}$ is the mass-transfer rate in units of 10$^{16}$ g s$^{-1}$ and $P_{hr}$ is the orbital period in hours. This equation assumes a weak dependence on the total mass of the system, $M$, and the mass of the white dwarf, $M_{1}$ (for more, see \citealt{Patterson2002}).

By changing the limit on the inner disk radius to $R_{d}\leq 0.05a$, Equation \ref{patterson_mag_mom} becomes
\begin{equation}
\label{patterson_mag_mom2}
\mu_{33} \leq 0.008 \dot{M}_{16}^{1/2}P_{hr}^{7/6}
\end{equation}
Assuming a mass-transfer rate of $2 \times 10^{17}$ g s$^{-1}$ (a typical value for IPs; \citealt{Patterson1994}; \citealt{Ballouz2009}) and using the orbital period of 3.94 hours gives an upper limit on the magnetic moment of $\mu \leq 1.2 \times 10^{32}$ gauss cm$^{3}$ for the WD in Lanning 386. This upper limit on $\mu$, when compared to Figure 16 of \cite{Patterson2002}, is below the $\mu$ of polars, but within the limits of the predicted $\mu$ for IPs. However, \cite{Patterson2002} show that IPs can have magnetic moments as high as $10^{34}$ gauss cm$^{3}$, suggesting the magnetic moment of Lanning 386 is on the low side for IPs. This, coupled with the low X-ray luminosity found in Section \ref{IsItMagnetic}, suggests Lanning 386 is very weakly magnetic, even for an IP. This is probably why the system has not been identified as an IP previously.

\subsection{The X-ray flux limit for J1923} \label{comp_both}
\cite{kennedy16} proposed that J1923 and Lanning 386 belong to the same CV class. Assuming comparable X-ray luminosities for J1923 and Lanning 386 (based off of their similar orbital periods and the similarities in their optical spectra and light curves), and using distances of 750$\pm$250pc \citep{kennedy16} and 160$\pm$50 pc respectively, the observed X-ray flux of Lanning 386 suggests that J1923 should be detected at a flux of $\sim5^{+5}_{-4}\times$10$^{-14}$ erg cm$^{-2}$ s$^{-1}$. However, we do not detect J1923, to a limiting flux of $1.1\times10^{-14}$ erg cm$^{-2}$ s$^{-1}$ at the 2$\sigma$ level. This is barely within the range of expected fluxes, and has 3 possible explanations. The first is that J1923 is at the far end of the distance limit. The second is that, due to the higher inclination in J1923 when compared to Lanning 386, the accretion disk in J1923 is blocking a significant proportion of the X-rays, lowering the flux. The third is that J1923 is intrinsically even less luminous in the X-rays than Lanning 386.

\section{Conclusion}
The requirement of a complex absorption model in the X-ray modelling of Lanning 386 favours a magnetic primary in the system, with the partial covering fraction and lower limit on the plasma temperature comparable to other IPs. The EW and fluxes from the Fe lines in Lanning 386 remain a bit of a mystery, as they do not match the expected values for a conventional IP. However, Lanning 386 best resembles the well studied IP/SW Sex system EX Hya. Additionally, the QPOs are possibly consistent with a magnetic primary. The limit on the magnetic moment of $\mu \leq 1.2 \times 10^{32}$ gauss cm$^{3}$, the low X-ray luminosity of $L_{X(2-10)}=\left(1.6^{+2.0}_{-0.9}\right)\times$10$^{30}$ erg s$^{-1}$ and the low ``hot'' plasma temperature of $kT=9^{+4}_{-2}$ keV suggest the magnetic field is very low when compared to other IPs.

However, there are still many questions surrounding systems like Lanning 386 and J1923. Lanning 386 lacks the strong persistent \ion{He}{ii} lines associated with magnetic accretion, as the \ion{He}{ii} are only strong when the system is in outburst. When in quiescence, as the system was for these observations, the \ion{He}{ii} lines are faint. There has also only been a tenuous detection of a periodicity in the X-ray light curve. Further X-ray observations are encouraged to test whether the 18.7 min period is real and coherent. It is also important to obtain an X-ray spectrum of Lanning 386 when in outburst as the dramatic differences seen between the outburst and quiescent optical spectrum may be reflected in the shape of X-ray spectrum or in the strength of its emission lines.

Finally, we didn't detect J1923 with an upper limit on the X-ray flux of $1.1\times10^{-14}$ erg cm$^{-2}$ s$^{-1}$ at the 2$\sigma$ level. This is still in line with J1923 and Lanning 385 having similar X-ray luminosities and further X-ray observations of J1923 to try and detect it are encouraged.

\section*{Acknowledgements}
We would like to thank the anonymous referee for their constructive feedback, which has improved the quality of this paper. MRK, PC and PMG acknowledge financial support from the Naughton Foundation, Science Foundation Ireland and the UCC Strategic Research Fund. MRK and PMG acknowledge support for program number 13427 which was provided by NASA through a grant from the Space Telescope Science Institute, which is operated by the Association of Universities for Research in Astronomy, Inc., under NASA contract NAS5-26555.  We thank the Vatican Observatory and Richard Boyle for providing us observing time on the VATT. This paper used data obtained with the MODS spectrographs built with funding from NSF grant AST-9987045 and the NSF Telescope System Instrumentation Program (TSIP), with additional funds from the Ohio Board of Regents and the Ohio State University Office of Research. The LBT is an international collaboration among institutions in the United States, Italy and Germany. The LBT Corporation partners are: The University of Arizona on behalf of the Arizona university system; Istituto Nazionale di Astrofisica, Italy;  LBT Beteiligungsgesellschaft, Germany, representing the Max Planck Society, the Astrophysical Institute Potsdam, and Heidelberg University; The Ohio State University; The Research Corporation, on behalf of The University of Notre Dame, University of Minnesota and University of Virginia. APO is a 3.5m telescope owned and operated by the Astrophysical Research Consortium. This research made use of Astropy, a community-developed core Python package for Astronomy \citep{2013A&A...558A..33A}. We would like to thank the anonymous referee for their insightul comments.

\bibliographystyle{mnras}
\bibliography{lanning_xmm.bib}

\label{lastpage}

\end{document}